\documentclass[preprint]{imsart}

\pdfoutput=1

\RequirePackage[OT1]{fontenc}
\RequirePackage{amsthm,amsmath, amssymb}
\RequirePackage[round, longnamesfirst]{natbib}
\RequirePackage[colorlinks,citecolor=blue,urlcolor=blue]{hyperref}
\usepackage{bbm}
\usepackage{subfigure}
\usepackage{graphicx}

\startlocaldefs
\numberwithin{equation}{section}
\theoremstyle{plain}

\endlocaldefs

\graphicspath{{./figures/}}

\begin{document}

\begin{frontmatter}
\title{Bayesian quantile regression analysis for continuous data with a discrete component at zero}
\runtitle{Bayesian quantile regression analysis for continuous data}

\begin{aug}
\author{\fnms{Bruno} \snm{Santos}\ead[label=e1]{bramos@ime.usp.br}}
\and
\author{\fnms{Heleno} \snm{Bolfarine}\ead[label=e2]{hbolfar@ime.usp.br}}

\address{Institute of Mathematics and Statistics\\
Rua do Mat\~ao 1010, Cidade Universit\'aria \\
S\~ao Paulo, Brazil \\
\printead{e1,e2}}



\runauthor{Santos and Bolfarine}

\affiliation{Universidade de S\~ao Paulo}

\end{aug}

\begin{abstract}
In this work we show a Bayesian quantile regression method to response variables with mixed discrete-continuous distribution with a point mass at zero, where these observations are believed to be left censored or true zeros. We combine the information provided by the quantile regression analysis to present a more complete description of the probability of being censored given that the observed value is equal to zero, while also studying the conditional quantiles of the continuous part. We build up an Markov Chain Monte Carlo method from related models in the literature to obtain samples from the posterior distribution. We demonstrate the suitability of the model to analyze this censoring probability with a simulated example and two applications with real data. The first is a well known dataset from the econometrics literature about women labor in Britain and the second considers the statistical analysis of expenditures with durable goods, considering information from Brazil.
\end{abstract}


\tableofcontents
\end{frontmatter}

\section{Introduction}

In the econometrics literature, a well known problem is the case when there is a non-negative continuous variable with a point mass at zero. \citet{tobin:58} considered the problem of expenditures of durable goods, assuming that all zero observations of expenditures of a household were actually censored observations. So in order to have a better understanding of the conditional mean of this response variable given the household income, for instance, a variable $Y^\star$ which is only observable when its value is positive is added in the modeling scheme. Then considering that the expenditures are normal distributed, one could use the normal cumulative distribution function in the likelihood computation. On the other hand, \citet{cragg:71} still investigating the effects of explanatory variables in durable goods purchase, defined a two-part model, where one part of the model studies whether the individual makes the purchase or not, then another part tries to explain how much is spent on average conditional on some variables. It is notable that these approaches deal with a similar problem using very different assumptions. Here in this paper we try to combine these ideas, without relying solely on the conditional mean to make inference about the continuous distribution, but instead we look for different quantiles of this distribution to accomplish this task.

Recently, there has been a surge of methods that give attention to other parts than the mean of the conditional distribution, often without trying to describe this distribution with just one family of distributions. One of these models is the quantile regression model, which was first proposed by \citet{koenker:78}, and is the one we consider in this work. Essentially, if one believes that the regression parameters are not fixed for the entire distribution, but rather depends on the quantile of interest, then these models are able to capture this effect. For example, these models are capable of measuring differences between central and tail estimates in the conditional distribution of the response variable, which is not usually the aim of conditional mean models. Some interesting applications of quantile regression models can be seen in \citet{yu:03}, \citet{koenker:05} and \citet{elsner:08}.

In the frequentist framework, the quantile regression parameters are obtained using linear programming algorithms, since the minimization problem proposed can be written as linear programming problem. For a Bayesian setting, the asymmetric Laplace distribution can be useful in obtaining posterior conditional quantile estimates. \citet{yu:01} proposed the use of this distribution in order to introduce a Bayesian quantile regression model. The authors verified in simulation studies that this assumption was helpful in approximating conditional quantiles for different probability distributions. \citet{yue:11} used this distribution to build additive mixed quantile regression models, where they adopted the Integrated Nested Laplace approximations (INLA) approach for posterior inference. Later, \citet{sriram:13} proved that this distribution is capable of providing good estimates, if certain conditions are satisfied. Moreover, \citet{kozumi:11} suggested a more efficient Markov Chain Monte Carlo (MCMC) method, making use of location-scale mixture representation of the asymmetric Laplace distribution, and even implementing a Tobit quantile regression when left censoring is present. This new approach made easy for others extensions in quantile regression models. For instance, \citet{lum:12} developed the asymmetric Laplace process to produce quantile estimates when the data presents spatial correlation. \citet{alhamzawi:12a} and \citet{alhamzawi:13} presented variable selection methods also considering this representation. \citet{luo:12} used this approach for longitudinal data models with random effects. 

Furthermore, \citet{santos:15} considering proportion data with inflation of zeros or ones developed a two-part model using a Bayesian quantile regression model to explain the conditional quantiles of the continuous part. They considered the equivariance property of the quantile function to work with a transformed variable in the modeling process to match the support of the asymmetric Laplace distribution. We aim to extend their model in this paper, but only concentrating in the zero inflation process, where we assume that a percentage of the zeros are left censored. By examining the continuous distribution with their conditional quantiles, we plan to convey more information about this probability of being censored given that the observation is zero. 

This paper is organized in the following manner. In Section 2, we show the Bayesian two-part model using quantile regression for the continuous part, with its prior and posterior setting. In Section 3, we extend the two-part model to allow that zero observations are either censored or are true zeros, defining the posterior probability of being censored given that it is zero. Advancing, in Section 4, we present the suitability of the method using a simulated example that compares the censoring probabilities for zero observations. Two applications of the model are presented to illustrate the results in Section 5. We complete with our final remarks in Section 6.

\section{Two-part model review}

If we consider the possibility of modeling the response variable by a mixture of two distributions, a point mass distribution at zero and a continuous distribution for the positive values, we can use the two-part model introduced by \citet{cragg:71}. Then we can write the probability density function of $Y$ as 
\begin{equation} \label{santos:eqDens}
 g(y) = p \mathbb{I}(y = 0) + (1-p)f(y) \mathbb{I}(y > 0),
\end{equation}
where $\mathbb{I}(a)$ is the indicator function, which is equal to one when $a$ is true and $p = P(Y = 0)$. For the example with durable goods, this model assumes that a person decides whether it is going to make a purchase or not in the first place and later one decides about how much it will spend. For each part of the process, the analysis can verify the importance of predictors to explain the probability of being zero and also in the outcome of the positive part. \citet{cragg:71} considered a probit link to study the probability $p$ and a truncated normal probability to fit the values greater than zero. With these assumptions, or even others distributions to the continuous part, it is possible to make inference about the conditional mean of the positive values. But we believe that it is important to check if other parts of the conditional distribution might present different conclusions about the association between the response variable and its explanatory variables. And in order to help this goal, we select to use quantile regression models.

\citet{santos:15} proposed the use of the asymmetric Laplace distribution for the continuous part, for proportion data, when the response variable is defined between zero and one. They used the equivariance property of the quantile function to model a transformed response to meet the limited support of proportion data to the real line support of the asymmetric Laplace distribution. And this distribution allows the posterior inference about the quantiles of the response variable, as proved by \citet{sriram:13}, proposed initially by \citet{yu:01} and later improved by \citet{kozumi:11}. Its probability density function, with location parameter $\mu$, scale parameter $\sigma$, skewness parameter $\tau$, can be written as
\begin{equation} \label{densALD}
 f(y | \mu, \sigma, \tau) = \frac{\tau(1-\tau)}{\sigma} \exp \left\{ - \rho_\tau \left( \frac{y - \mu}{\sigma} \right) \right\},
\end{equation}
where $\rho_\tau(u) = u(\tau - \mathbb{I}(u < 0))$, $\mu \in \mathbb{R}$ is the $\tau$th quantile, $\sigma > 0$ and $\tau \in [0,1]$. With the intention to produce information about the conditional quantile given some predictors, for a regression analysis purpose, we write the location parameter with a linear predictor, i.e., $\mu = x^\prime \beta(\tau)$, where the regression parameter $\beta(\tau)$ is indexed by $\tau$ to indicate its connection with the quantile of interest, as $\tau$ is usually fixed during the analysis. We consider here a location-scale representation mixture of the asymmetric Laplace distribution, which was used by \citet{kozumi:11} to present a more efficient Gibbs sampler for quantile regression models. We can say that if $Y$ is distributed according to an asymmetric Laplace distribution with parameters $\{\mu, \sigma, \tau\}$, then we have
\begin{align*} \label{mixtureRep}
 Y | v &\sim N(\mu + \theta v, \psi^2 \sigma v) \\
 v &\sim \mbox{Exp}(\sigma)
\end{align*}
where 
\begin{equation*}
 \theta = \frac{1-2\tau}{\tau(1-\tau)}, \quad \quad  \psi^2 = \frac{2}{\tau(1-\tau)}.
\end{equation*}

We can also add covariates to explain the probability of being equal to zero for each observation, $p_i$, by making $p_i = \eta(z_i^\prime \gamma)$, where $\eta(.)$ is a link function, that could be, for instance, the normal cumulative distribution function (cdf), producing the probit model or the logistic cdf, producing the logistic model. The set of variables in this case can be either different or the same as the one used to describe the continuous density.

If we define the sets $J = \{ y_i : y_i = 0 \}$, and $K = \{ y_i : y_i > 0\}$, the augmented likelihood function for the two-part model considering the location-scale mixture of the asymmetric Laplace distribution, can be defined as
\begin{equation} \label{twopartLik}
 L(\beta(\tau), \gamma, \sigma) = \prod_{y_i \in J} \eta^{-1}(z_i^\prime \gamma) \prod_{y_i \in K} \left(1-\eta^{-1}(z_i^\prime \gamma)\right) f(y_i | v_i) f(v_i).
\end{equation}

To complete the Bayesian specification, we assume priors distributions for the parameters, with a normal distribution for $\gamma$ and $\beta(\tau)$, and an inverse gamma distribution for $\sigma$. Given these definitions we can write the full hierarchical model as
\begin{align*}
 Y_i | v_i &\sim p_i I(y_i \in J) + (1-p_i) N(x_i^\prime \beta(\tau) + \theta v_i, \psi^2 \sigma v_i) I(y_i \in K), \\
 v_i &\sim \mathcal{E}(\sigma), \\
 p_i &=  \eta(z_i^\prime \gamma), \\
 \beta(\tau) &\sim N(b_0, B_0), \\
 \sigma &\sim \mathcal{IG}(n_0, s_0), \\
 \gamma &\sim N(g_0, G_0). \\
\end{align*}

The full conditional distributions for all parameters, after we combine the likelihood with the prior information, are
\begin{align*}
\beta(\tau) \mid y,  v, \gamma, \sigma &\sim N(b_1, B_1), \\
v_i \mid h(y), \beta(\tau), \gamma, \sigma &\sim \mathcal{GIG}(1/2, \hat{\delta}_i, \hat{\xi}_i), \\
\sigma \mid y, v, \beta(\tau), \gamma &\sim \mathcal{IG}(\tilde{n}/2, \tilde{s}/2), \\
 \pi(\gamma \mid y, v, \beta(\tau), \sigma) &\propto \prod_{i \in C} \eta^{-1}(z_i^\prime \gamma) \prod_{i \in D} \left(1-\eta^{-1}(z_i^\prime \gamma)\right)
\exp \left\{- \frac{1}{2} (\gamma - g_0)^\prime G_0^{-1} (\gamma - g_0) \right\}. \\
\end{align*}

With the exception of $\gamma$, all parameters of the full conditional distributions are similar to the ones in the MCMC proposed by \citet{kozumi:11}, and for that reason are not provided here. $\mathcal{GIG}(\nu, \delta, \zeta)$ represents a generalized inverse Gaussian distribution, for which we can use the algorithm by \citet{dagpunar:89} to generate values from the posterior distribution for each $v_i$.

The posterior distribution for $\gamma$ has not a recognizable distribution, so we suggest a random walk Metropolis-Hastings algorithm, where we can use as proposal a multivariate normal distribution centered at the current value of $\gamma$. Then at the $k$th step of the MCMC, we draw $\gamma^{(k)}$ from $N(\gamma^{(k-1)}, \sigma^2_\gamma \Omega_\gamma)$, and $\gamma^{(k)}$ is accepted with probability
\[
 \alpha(\gamma^{(k)}, \gamma^{(k-1)}) = \min \left\{ 1, \frac{\pi(\gamma^{(k)} \mid y, v, \beta(\tau), \sigma)}{\pi(\gamma^{(k-1)} \mid y, v, \beta(\tau), \sigma)} \right\}, 
\]
where $\sigma^2_\gamma$ is a tuning parameter that should be chosen carefully to give acceptance probabilities between 0.15 and 0.50 \citep[see][]{gelman:03}. We define $\Omega_\gamma$ as the identity matrix, as this choice has given us adequate results in both simulated data and applications, but other options could be studied. All other full conditional distributions are relatively easy to generate posterior samples. Upon request, we can share the code we used to implement this algorithm, which was written in R.

\section{Bayesian quantile regression for continuous data with a point mass distribution at zero}

Going back to the problem of studying expenditures in durable goods in a determined period, we have that a proportion of households might not have made any purchase for these kinds of items, having zero as the total of expenditures. This result was analyzed using two different approaches in \citet{tobin:58} and \citet{cragg:71}. Here we aim to combine these ideas, while we make inference about the conditional quantiles of the continuous element of this type of problem. In the biometrics literature, \citet{moulton:95} proposed an extension of Cragg's model, considering that an observed  zero, or a lower detection limit, can be either from the point mass distribution or from the continuous distribution, being in the latter case a censored observation. \citet{chai:08} considered a similar formulation, but changing the distribution of the continuous part and also the link function to model the probability $p_i$. In the survival analysis literature, an analogous situation occurs when there is a proportion of the sample which is believed to be cured from a possible illness. As a consequence, at the end of study from all the observations that are not affected by the event of interest, some are assumed to be cured and others are considered censored observations.

In this case, we should rewrite the density in \eqref{santos:eqDens} as
\begin{equation} \label{mixtCens}
 g(y) = [p + (1-p)F(0)]\mathbb{I}(y = 0) + (1-p)f(y) \mathbb{I}(y > 0)
\end{equation}
where $F(.)$ is the cdf of the continuous part. In our quantile regression models, for the censored observations, we consider the idea proposed by \citet{chib:92} and adapted for quantile regression by \citet{kozumi:11}, which samples $Y*$, the assumed latent variable using 
\begin{equation} \label{truncProc}
 Y^\star \sim TN_{(-\infty, 0]}(\mu + \theta v, \psi^2 \sigma v),
\end{equation}
where $TN_{[a,b]}(\mu,\sigma^2)$ denotes a truncated normal distribution, with mean $\mu$ and variance $\sigma^2$, in the interval $[a,b]$. If we define the variable $C$ as a censoring indicator, where $C$ is equal to one when $Y$ is censored and zero otherwise, we have that the probability of being censored given that $Y$ is zero is
\begin{equation} \label{probCens}
 P(C = 1 | Y = 0) = \frac{(1-p)F(0)}{p + (1-p)F(0)},
\end{equation}
where $F(0)$ depends on the parameters of the asymmetric Laplace distribution and can be defined as 
\begin{equation*}
 F(0; \mu, \sigma, \tau) = \begin{cases}
         \tau \exp \left\{ - \frac{(1-\tau) \mu}{\sigma} \right\}, & \mbox{ if }  \mu \geqslant 0, \\
         1 - (1-\tau) \exp \left\{\frac{\tau \mu}{\sigma} \right\}, & \mbox{ if }  \mu < 0.
        \end{cases}
\end{equation*}

\begin{figure}
\begin{center}
\subfigure[]{
\resizebox*{5.5cm}{!}{\includegraphics{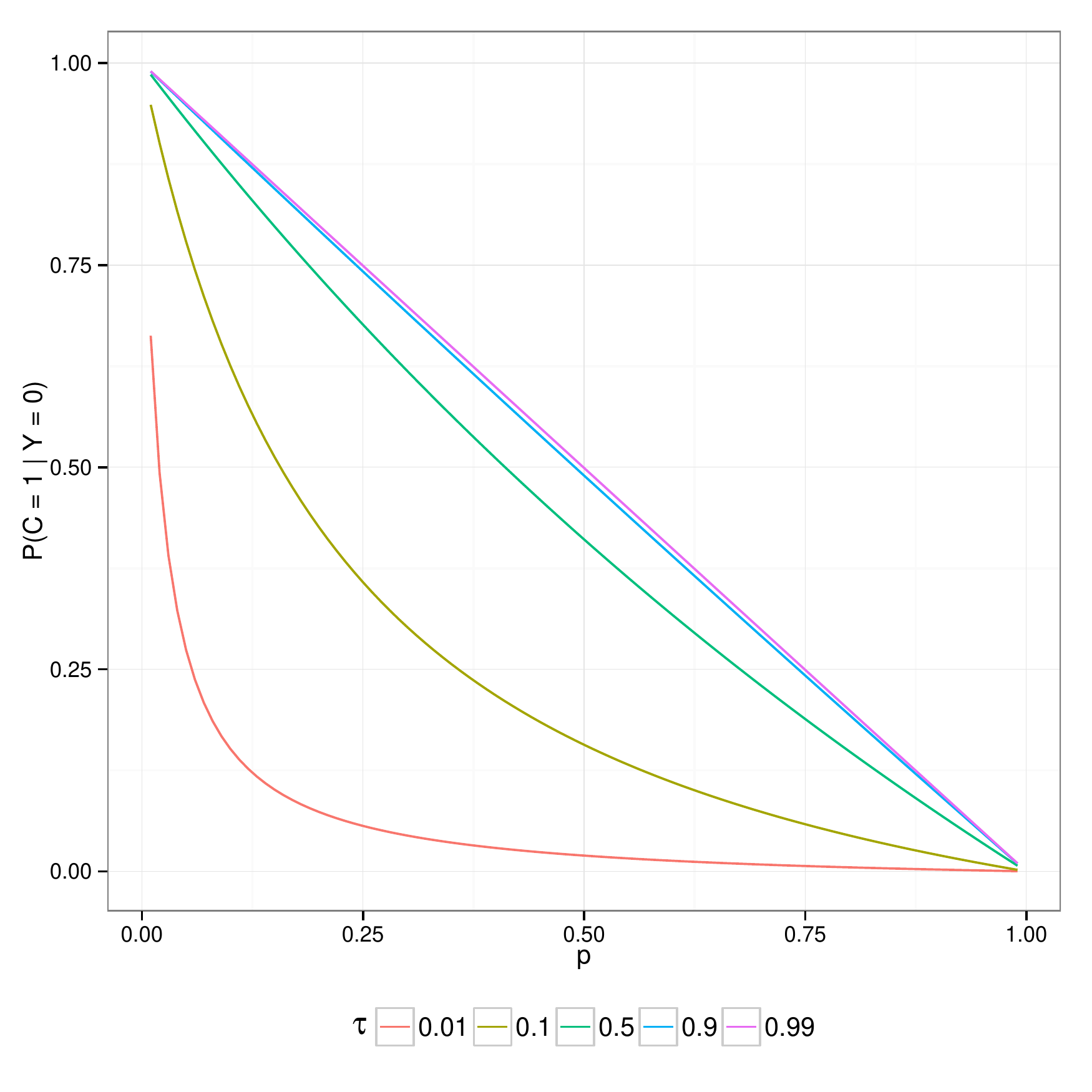}}}\hspace{5pt}
\subfigure[]{
\resizebox*{5.5cm}{!}{\includegraphics{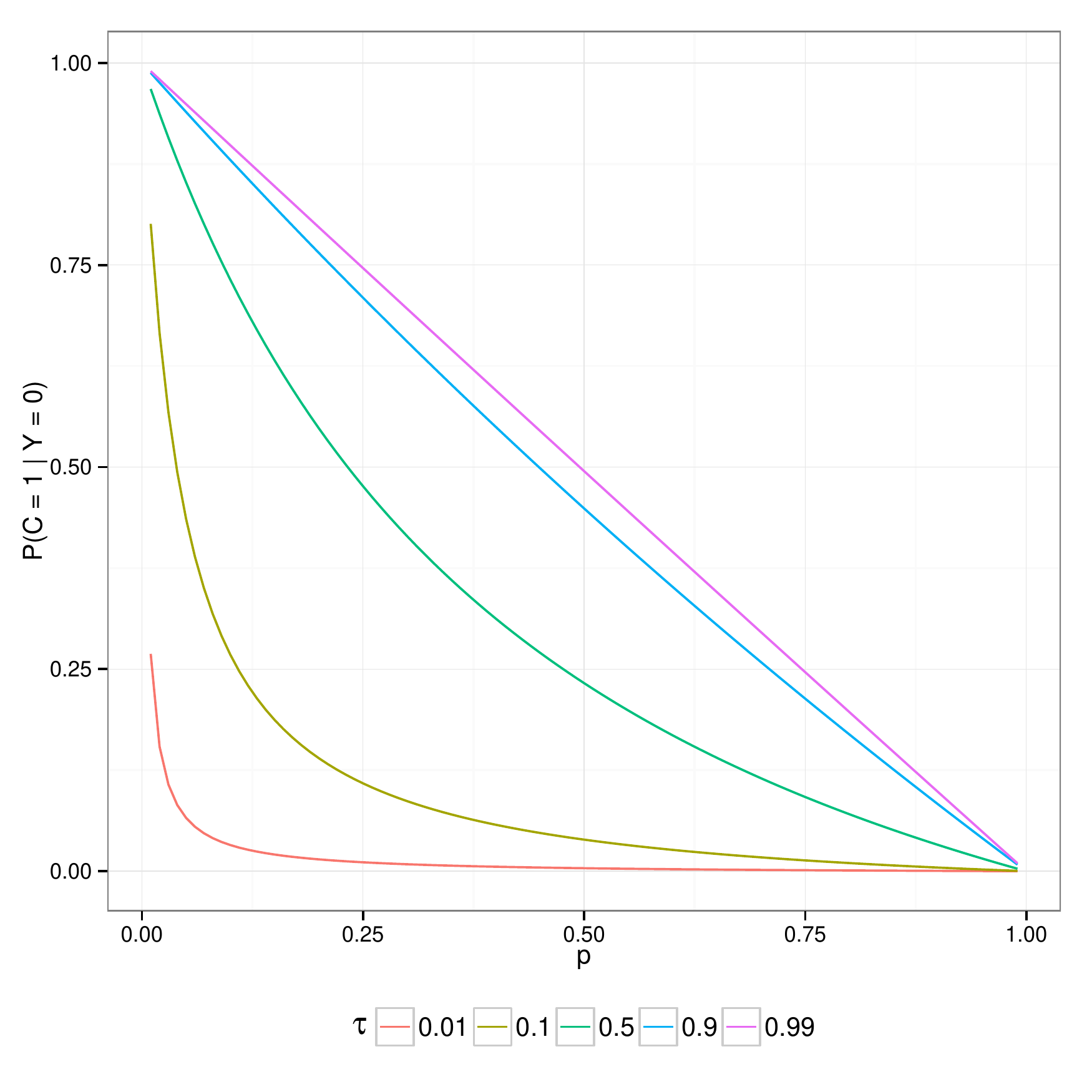}}}
\caption{\label{fig1} Plots for the probability of being censored for different $\tau$'s as a function of the probability $p = P(Y = 0)$. (a) $x^\prime \beta(\tau) = -1$, (b) $x^\prime \beta(\tau) = 1$}
\label{figure1}
\end{center}
\end{figure}

If we calculate the probability in \eqref{probCens} when $\sigma = 1$, and for two possible values for $\mu = x^\prime (\beta)$, -1 and 1, considering different values for $p$, we can draw the curves in Figure~\ref{figure1}. Comparing the two possible values, we have the probability is being censored is greater when the linear predictor is equal to -1, for the same $\tau$. It is easy to show that this probability is limited between zero and $1-p$ and this characteristic is depicted in the plot. Also, for the same $p$ this is probability is increasing in $\tau$, which is controlled by $F(0)$. 

This latent variable, $C$, which indicates the censoring mechanism is non-observable for all zero observations. Therefore, we need to use a data augmentation algorithm in this case. Our complete cases are $\{Y_i, C_i, v_i\}$, from which we only observe $Y_i$. We showed in the previous section how to update $v_i$, which is a necessary feature in the location-scale mixture of the asymmetric Laplace distribution. In order to update the censoring indicator $C_i$ we use the probability in \eqref{probCens}, which will depend on the current values of the parameter related to the probability $p$, but also to the parameters related to the continuous part, namely the quantile regression parameters. This interesting property grants that a certain observation, with its response value equal to zero, can have different probabilities of being censored given its conditional quantile estimate and its conditional probability of being equal to zero. For the complete cases, define the sets $C = \{ y_i : y_i = 0 \mbox{ and } c_i = 1\}$, $D = \{ y_i : y_i = 0 \mbox{ and } c_i = 0\}$, and $K = \{ y_i : y_i > 0\}$, of censored observations, non-censored but with response equal to zero, and observations greater than zero, respectively. Then the likelihood function for $\xi = (\beta(\tau), \gamma, \sigma)$, without writing the conditional parameters for $F(0)$ and $f(y_i|v_i)$ for notational simplicity, can be written as 
\begin{equation*}
 L(\xi) = \prod_{y_i \in D} \eta^{-1}(z_i^\prime \gamma) \prod_{y_i \in C} (1 - \eta^{-1}(z_i^\prime \gamma)) F(0) \prod_{y_i \in K} \left(1-\eta^{-1}(z_i^\prime \gamma)\right) f(y_i | v_i) f(v_i).
\end{equation*}

It is important to note that $F(0)$ varies for every observation in $C$, given their linear predictor for the conditional quantile. Although, instead of evaluating $F(0)$ for every censored observation, we can replace the censored observation for its estimate based on \eqref{truncProc}. By doing this, our likelihood function resemble the likelihood for the two-part model of the previous section, and the MCMC described there can be used to update the parameters here as well.

A posterior estimate of the probability of being censored for each observation can be calculated as
\begin{equation} \label{eqProbCens}
 P(C_i = 1 | Y, v, \beta(\tau), \sigma, \gamma) = \sum_{k = b + 1}^M \frac{C_i^{(k)}}{M - b}, \quad \quad \forall i : y_i \in C \cup D, 
\end{equation}
where $C_i^{(k)}$ is the $k$th term of the Markov chain for the censoring indicator of the $i$th observation, $M$ is the length of the chain and $b$ is the length of the burn-in period.

\section{Simulation study for the censoring probability}

In this section, we are concerned in checking the performance of our model regarding its capability of making statements about the probability of being censored given that a certain observation has its response value equal to zero. In order to accomplish that, we replicate a study where we know which observations are censored and also which ones are not censored, between those with zero as their response value, and we compute this probability of interest for each group.

We consider a model with just two covariates and the following structure as
\begin{align*}
 \log \left( \frac{p_i}{1-p_i} \right) &= \gamma_0 + \gamma_1 z_{i1} + \gamma_2 z_{i2}, \\
 Y_i &= \beta_0 + \beta_1 x_{i1} + \beta_2 x_{i2} + \epsilon_i,
\end{align*}
where $\epsilon_i \sim N(0, 0.5^2)$, $\beta_0 = -0.5$, $\beta_1 = 0$ and $\beta_2 = 1.5$. We draw $x_{ij}$, $j=1,2,$ from a uniform distribution, and we set $x_{ij} = z_{ij}$, i.e., we use the same covariates for both parts of the model. We begin our study trying to create a scenario where there is a distinct difference between the observations that belong to the point mass distribution and the observations from the continuous part. Therefore, to achieve this goal we use big absolute values for $\gamma_1$ and $\gamma_2$, initially, 10 and -10, respectively. By defining these values, our intention is to give greater probability $p_i$ for observations which are true zeros. On average, 50\% of the sample is classified as a true zero in the start and then another 10\% is classified as zero after being censored. Our sample size in this study is 500 and we report our results for 1000 replications of this model. For the prior hyperparameters, we assumed the $b_0 = g_0 = 0$ and $B_0 = G_0 = 100I$, where $I$ is the identity matrix, and for $\sigma$ we considered $\mathcal{IG}(3/2, 0.1/2)$, as in \cite{kozumi:11}. All the results were based after discarding the first 500 posterior samples and considering the next 1500 draws from the posterior, calculating the posterior mean for each parameter in these draws.

For each simulation, we calculate the probability of being censored for all observations with $y_i=0$. Then we summarize all probabilities with the mean for the group of censored observations and also the group of non-censored observations, respectively as 
\begin{equation*}
 \zeta_C = \sum_{y_i \in C} \frac{P(C_i = 1 | Y_i = 0)}{n_C} \quad \quad \zeta_D = \sum_{y_i \in D} \frac{P(C_i = 1 | Y_i = 0)}{n_D},
\end{equation*}
where $n_C$ is the number of censored observations with $y_i=0$ and $n_D$ is the number of observations non-censored with $y_i = 0$, and $P(C_i = 1|Y_i = 0)$ is calculated according to \eqref{eqProbCens}.

\begin{figure}[!tb]
\begin{center}
\includegraphics[scale=0.35]{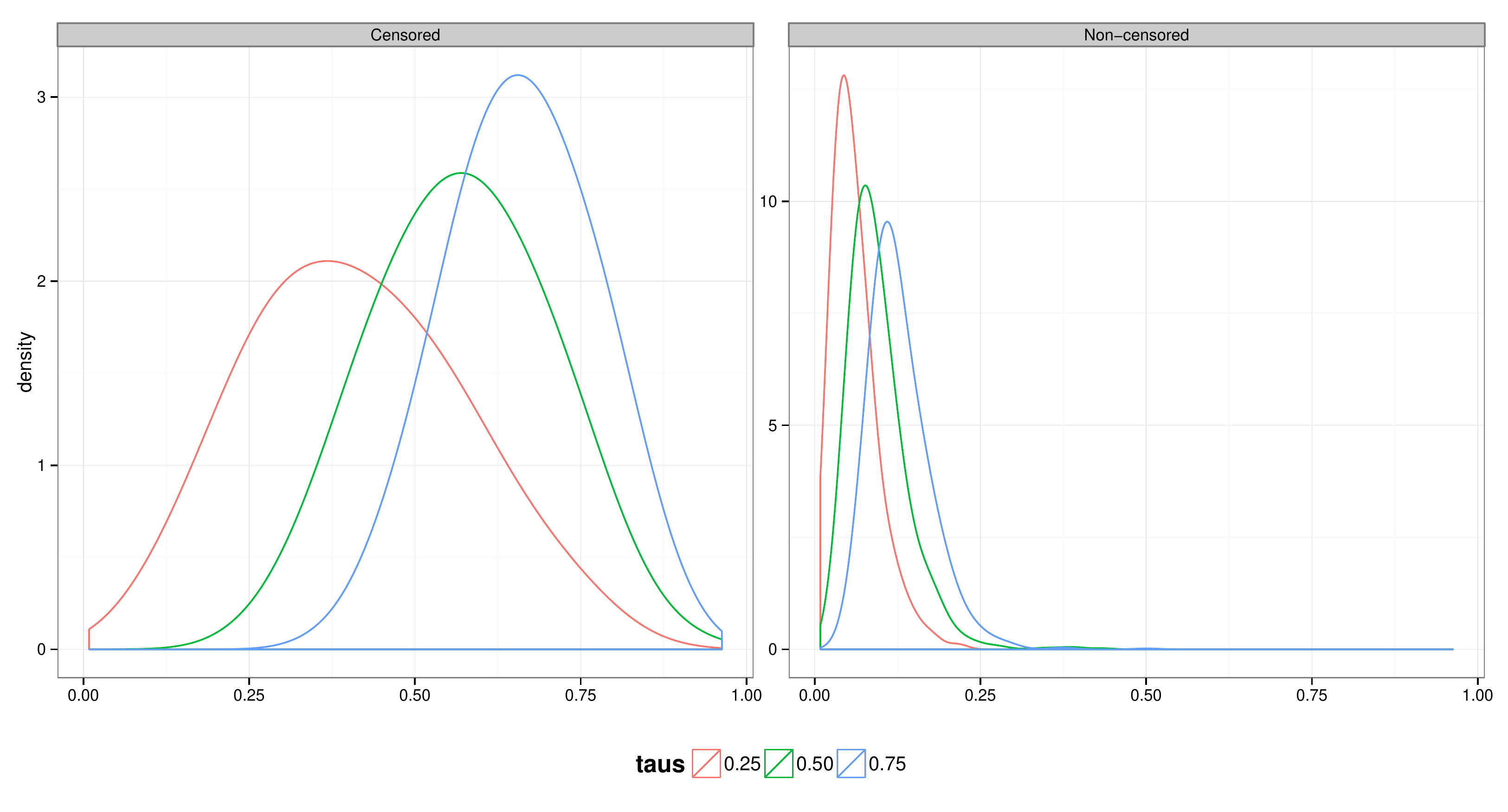}
\caption{Estimated density for the mean posterior probabilities of being censored for censored observations and also non-censored observations.}
\label{figure2}
\end{center}
\end{figure}

In Figure~\ref{figure2} we plot the estimated density for the 1000 $\zeta_C$ and $\zeta_D$ obtained, considering three different $\tau$'s: 0.25, 0.50, 0.75. We can see that for non-censored observations the density is mostly concentrated before 0.25, and that for this concentration is even more acute for the lower quantiles. For the censored group, we see that the probabilities of being censored are definitely larger than those for the non-censored, even though for the conditional quantile 0.25 is mostly concentrated around 0.30, but varying widely. For the conditional quantile 0.75, this probability is mainly concentrated around 0.70, therefore giving, on average, greater probability to censored observations in this case. Overall, we note that probabilities of being censored increase with the conditional quantile of interest, which was already discussed in Section 3.

Besides the interest in estimation of the probability of being censored, it is important to check whether the uncertainty about whether zero observations are censored or not undermine the estimation process of other parameters. There is just one note about the length of the MCMC chains in this simulation example, as we acknowledge that these sample sizes are small for the algorithm with a Metropolis-Hasting step, but we believe that this compromise was necessary due to time constraints in order to get some results in this simulation study.

\begin{figure}[tb]
\begin{center}
\includegraphics[scale=0.35]{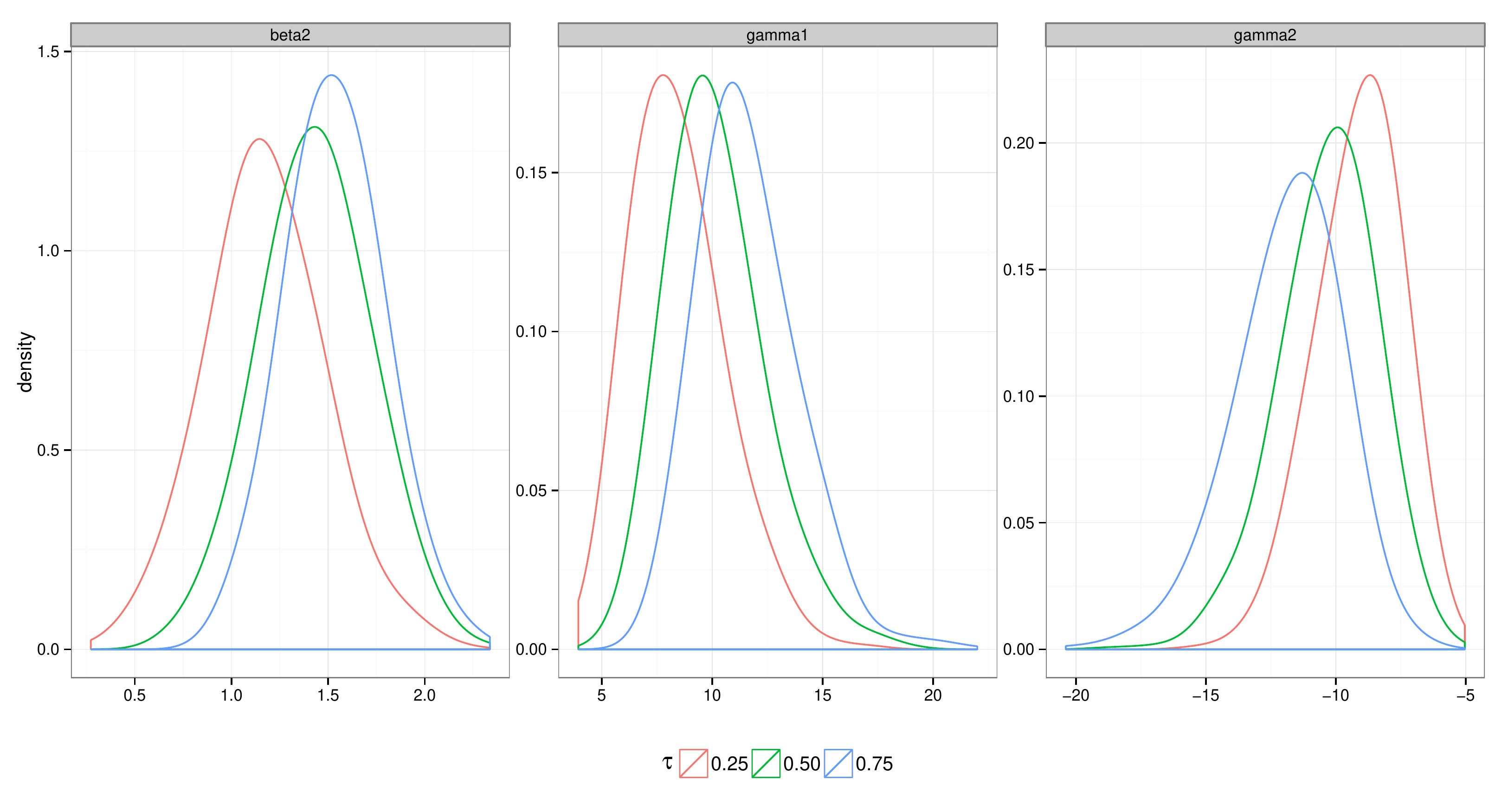}
\caption{Estimated density for the estimates of parameters $\beta_2$, $\gamma_1$, $\gamma_2$.}
\label{figure3}
\end{center}
\end{figure}

The density of the 1000 posterior estimates of $\gamma_1$, $\gamma_2$ and $\beta_2$ is depicted in Figure~\ref{figure3}. We are able to check that for all three quantiles, the estimation of these parameters is not affected by the uncertainty of the censoring process. In general, these parameters are reasonably estimated. We note, however, some interesting results about the variance of these estimates. Due to the smaller error of the censoring process for $\tau = 0.75$, since the mean probability of being censored is greater for this quantile, then $\beta_2$ is better estimated, with mean value closer to 1.5 and also with less variation around this value. On the other hand, the estimates for $\gamma_2$ show a similar performance, but in the 0.25 quantile. This is probably due to the fact that, in this quantile, the observations which are true zero get greater estimates for $p_i$. Therefore in the iteration  process the latent variable $C_i$, which is responsible to separate between the censored and non-censored observations,  it will be updated as 1 in fewer opportunities. Nevertheless, this interesting feature is less highlighted in the estimates of $\gamma_1$.

Ultimately, we do not add another distribution for the error in this study, instead of the normal distribution, as we believe that such a change would not provide more information about the effectiveness of our model. Moreover, considering a larger sample would be helpful, as we expect that estimate errors would decrease with larger samples, but since the results were already satisfactory for this sample size, we decided not to continue any further. And adding more variables to each part of the model would definitely make more difficult the estimation process, but then we understand that this is a complication with which the algorithms could deal separately.

\section{Applications}

We exemplify our model with two applications. First, we consider the data from \citet{mroz:87}, which was used for illustration purposes of the Tobit quantile regression model in \citet{kozumi:11}. And second, we present data about expenditures with durable goods in Brazil between 2008 and 2009, motivated by earlier considerations about this type of data by \citet{tobin:58} and \citet{cragg:71}.

\subsection{Labour supply data}

Analyzing empirical models about female labour supply, \citet{mroz:87} collected data about 753 married women, aged between 30 and 60 years old. The response variable of interest here is the number of hours worked for pay during the year of 1975, measured in 100h. In the sample, which was collected from the ``Panel Study of Income Dynamics'', there are 325 women who did not work in that year, so their response variable is equal to zero. In \citet{kozumi:11} they used this dataset to exhibit the Tobit quantile regression model, where these zero observations are assumed to be left censored. In our model, we combine the probability of being censored with the probability of being equal to zero to provide a more comprehensive study of these women who did not work that year. For covariates, we select non-wife income ({\it $x_1$}), years of education ({\it $x_2$}), actual years of labour market experience ({\it $x_3$}), wife's age ({\it $x_4$}), the number of children 5 years old or younger ({\it $x_5$}), and the number of children between 6 and 18 years old ({\it $x_6$}). After standardizing all covariates, we consider the following models for the probability and the conditional quantiles
\begin{align*}
 \log \left( \frac{p_i}{1-p_i} \right) &= \gamma_0 + \gamma_1 x_{i1} + \gamma_2 x_{i2} + \gamma_3 x_{i3} + \gamma_4 x_{i4} + \gamma_5 x_{i5} + \gamma_6 x_{i6}, \\
 Q_{Y_i}(\tau | x_i) &= \beta_0 + \beta_1 x_{i1} + \beta_2 x_{i2} + \beta_3 x_{i3} + \beta_4 x_{i4} + \beta_5 x_{i5} + \beta_6 x_{i6}.
\end{align*}

\begin{figure}
\begin{center}
\includegraphics[scale=0.40]{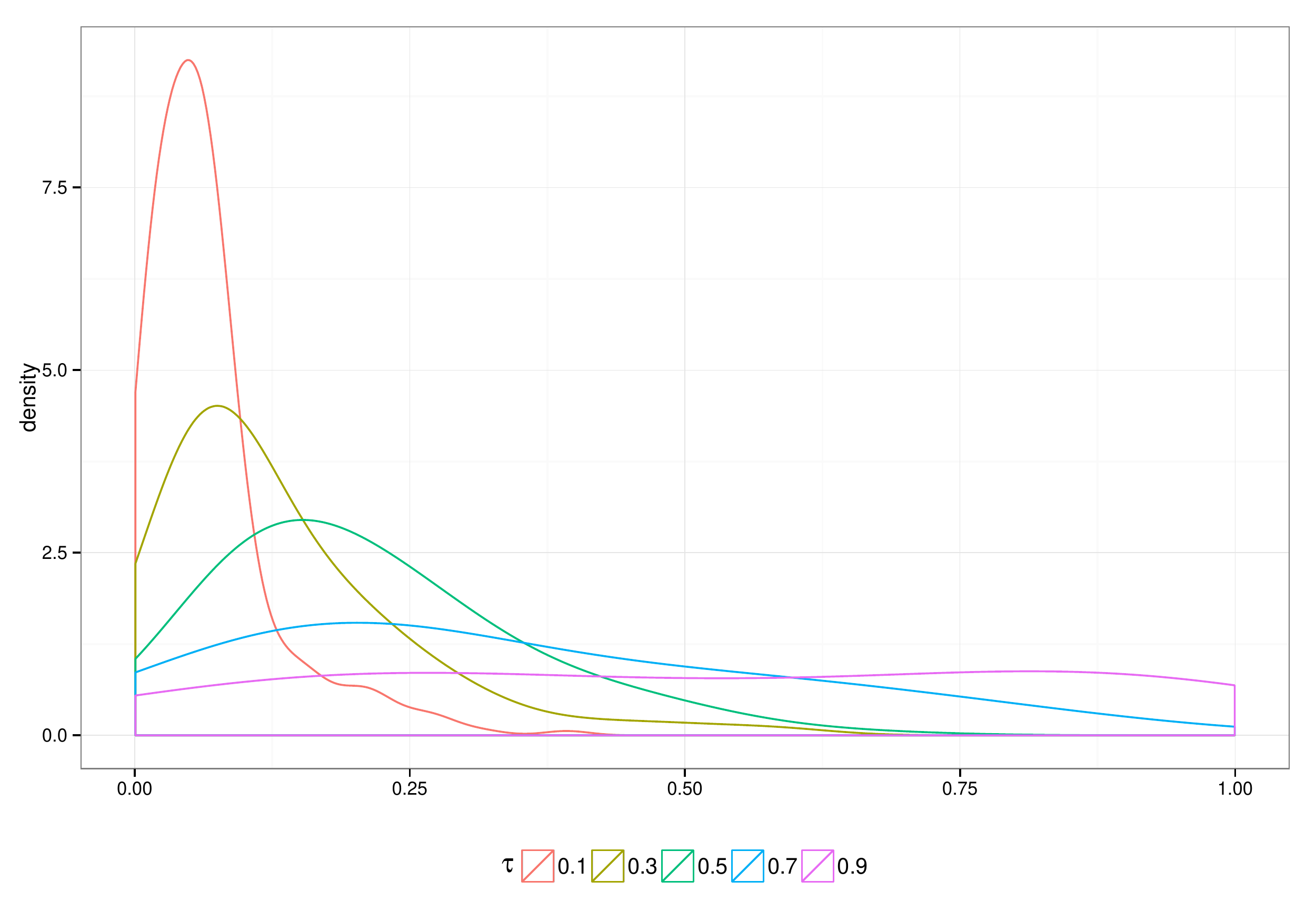}
\caption{Densities of the probability of being censored for $\tau = 0.1, 0.3, 0.5, 0.7, 0.9$.}
\label{figure4}
\end{center}
\end{figure}

\begin{figure}
\begin{center}
\includegraphics[scale=0.35]{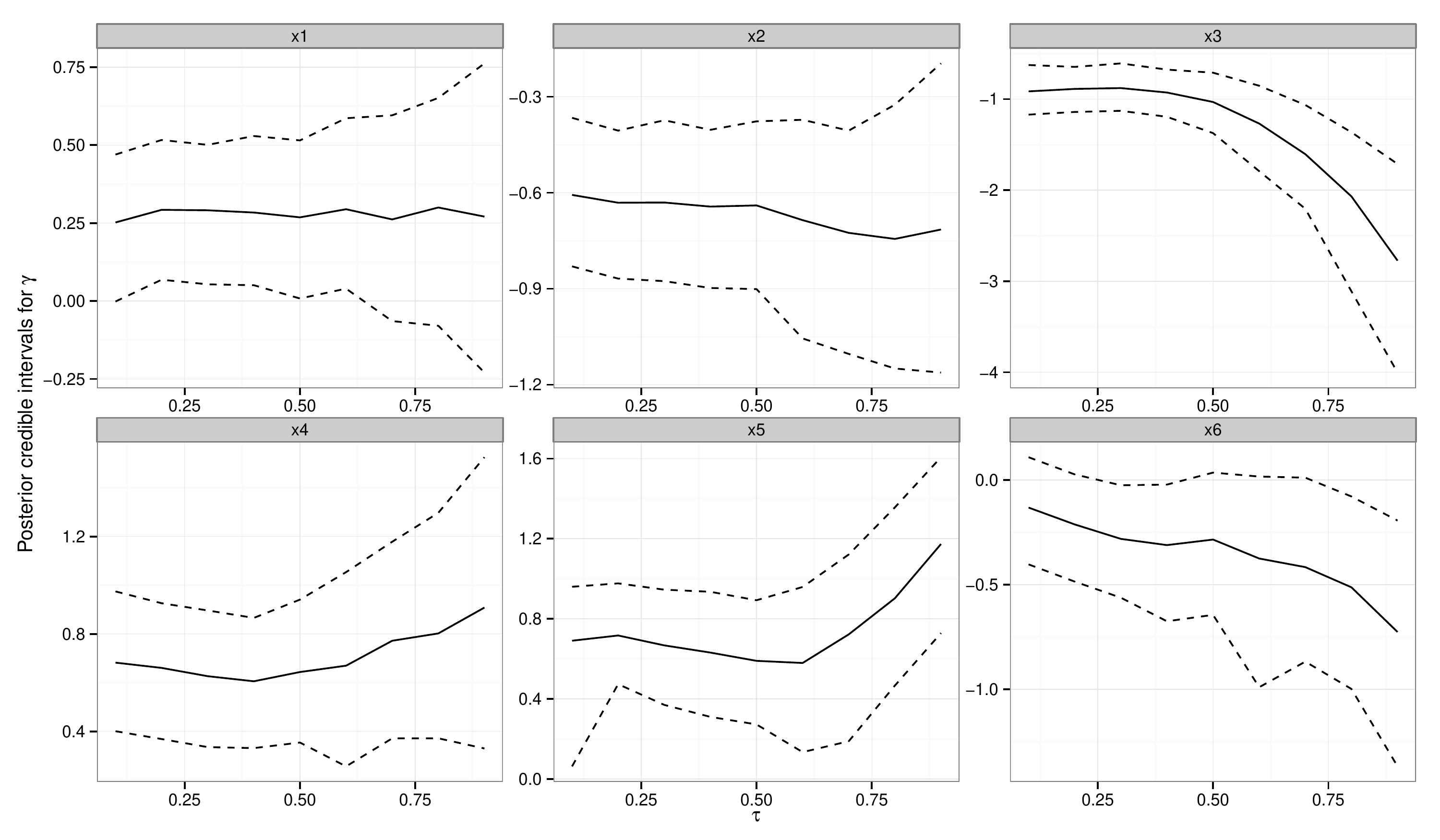}
\caption{Posterior mean and 95\% credible interval for $\gamma_i$, $i=1,\ldots,6$.}
\label{figure4_0}
\end{center}
\end{figure}

In Figure~\ref{figure4}, we present the densities of the probabilities of being censored given that a certain observation is equal to zero for distinct $\tau$'s. As mentioned earlier, these probabilities are $\tau$-dependent, therefore it is expected the very contrasting shapes of these densities. We estimate that for $\tau = 0.10$, these probabilities are mostly concentrated under 0.25, while for $\tau = 0.90$ they are well spread between 0 and 1, with mean value equal to 0.53. Having those differences between quantiles in the probabilities of being censored generates some variation in the posterior credible intervals for a few $\gamma$'s, as we can see in Figure~\ref{figure4_0}. For instance, for $\gamma_3$ there is a significant increase in absolute value for greater $\tau$. Also, the effect of the number of children aged older than 6 is estimated to be different than zero and negative just for greater quantiles, $\tau = 0.8$ and $\tau = 0.9$, while the coefficient for non-wife income could be considered significant just for lower quantiles. Overall, the coefficients for $x_1$, $x_4$ and $x_5$, when significant, are estimated to be positive. On the other hand, the estimates for the other variables are estimated negative to explain the probability of the hours of work being equal to zero. At this point, we should also index $\gamma$ by $\tau$ as well, but we let this feature just for $\beta(\tau)$. 

Besides that, we are able to compare the variation for each variable in the model according to the probability of being censored, which is a new factor that we add with our model for this type of analysis. For example, taking into account the variables income which is not due to the wife and years of experience, there are interesting results when we compare groups with different values for this probability for two values of $\tau$. 

\begin{figure}
\begin{center}
\includegraphics[scale=0.3]{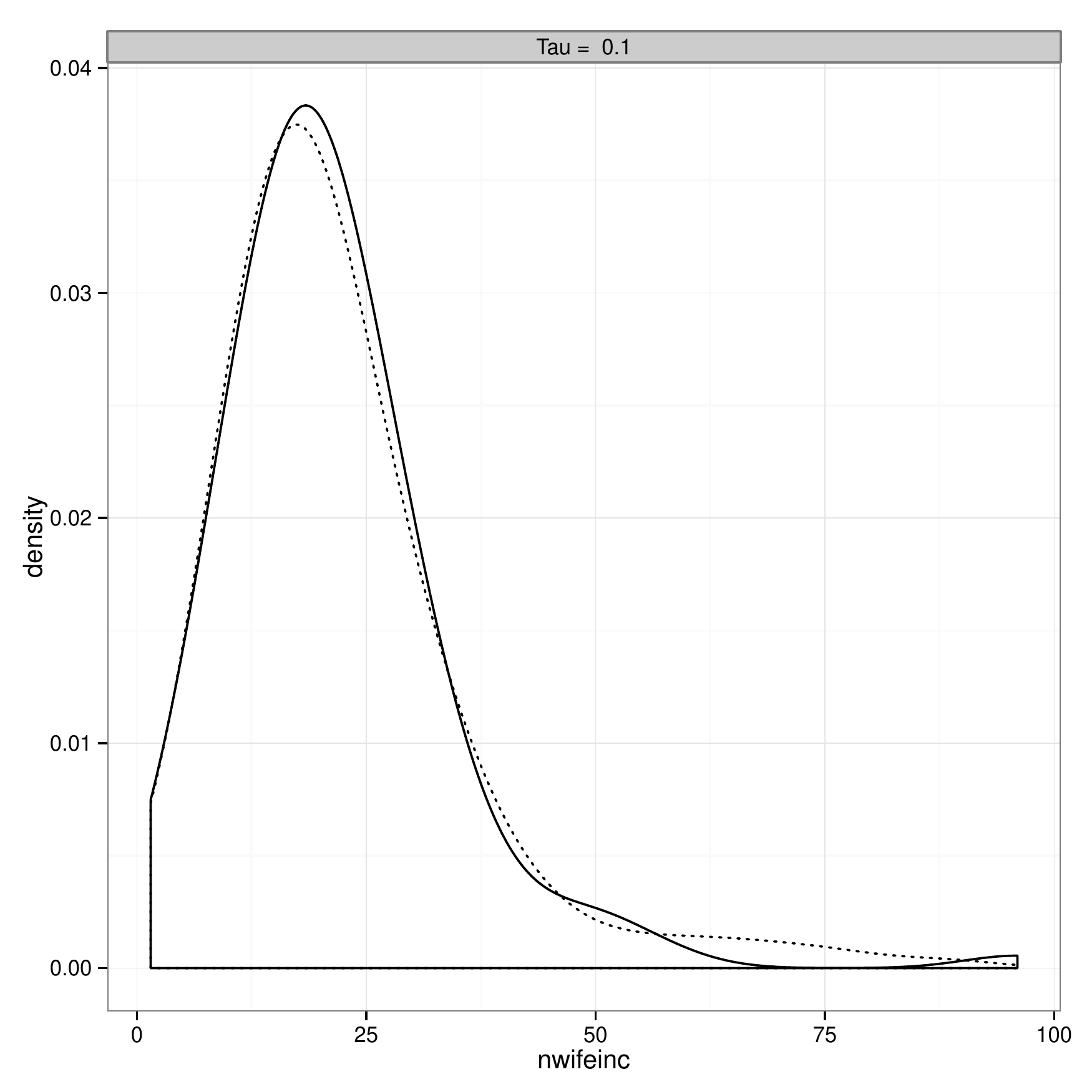}
\includegraphics[scale=0.3]{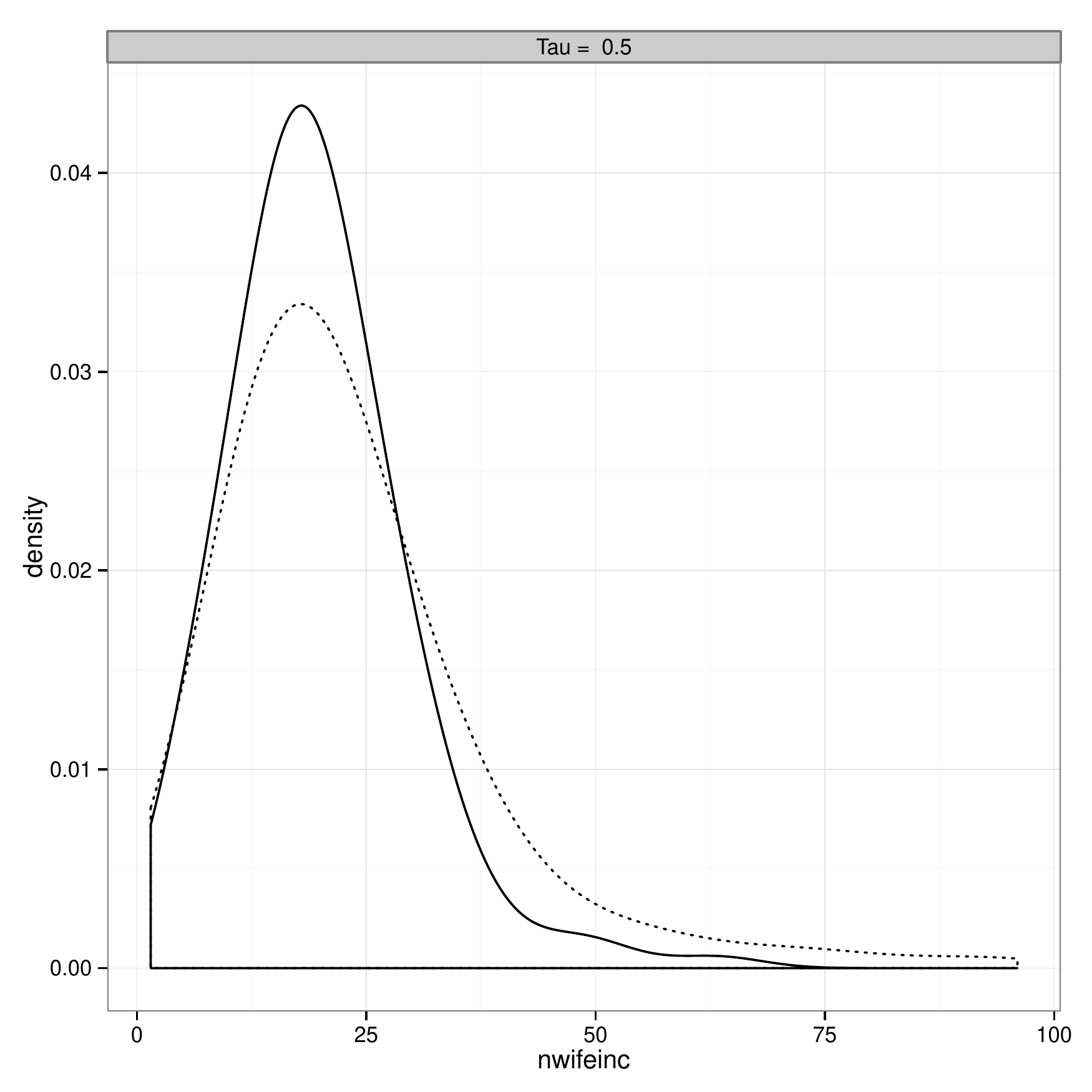}
\caption{Density of the variable {\it nwifeinc} separated in two groups according to their relative censoring probability in comparison with the mean probability in a given quantile, for $\tau = 0.1, 0.5$. Solid lines are for the group above the mean probability of being censored, while dashed lines are for the group below.}
\label{figure5}
\end{center}
\end{figure}

In Figure~\ref{figure5} we oppose the distribution of non-wife's income for women who have probability of being censored below and above the average for a specific quantile. For $\tau = 0.10$, the mean probability is equal to 0.07 and for $\tau = 0.50$ is 0.21. While there is not a conceivable difference in distribution in the quantile 0.10 between those two groups, for $\tau = 0.50$ there is a noticeable change in the distribution for women with lower than average probability of being censored against women with higher probability. Those women more inclined to be classified as true zeros, with lower than average probability of being censored, have a distribution of income more fat in the upper tail. We should note that this income, which is not due to the wife, could be seen as a greater incentive to work zero hours.

\begin{figure}
\begin{center}
\includegraphics[scale=0.3]{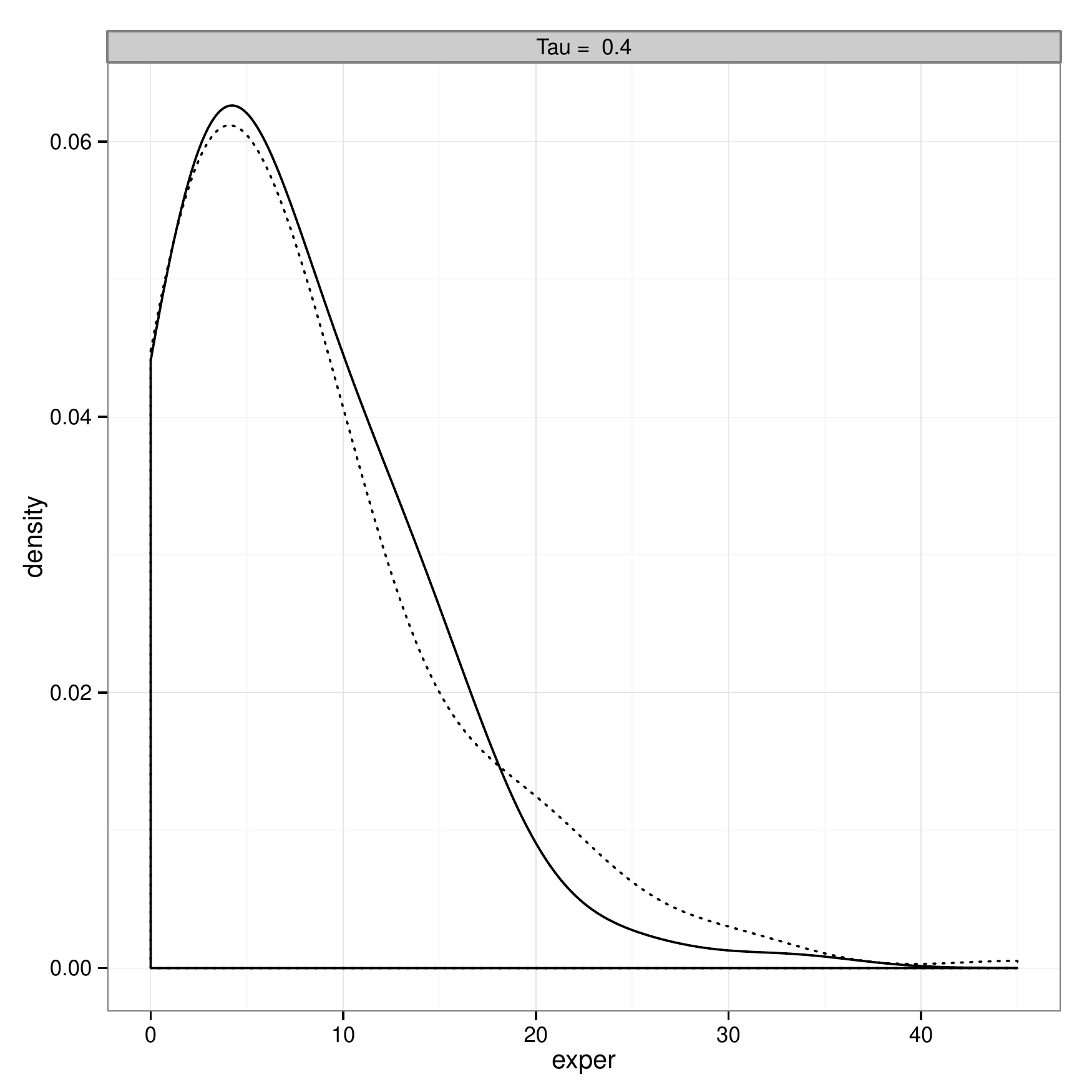}
\includegraphics[scale=0.3]{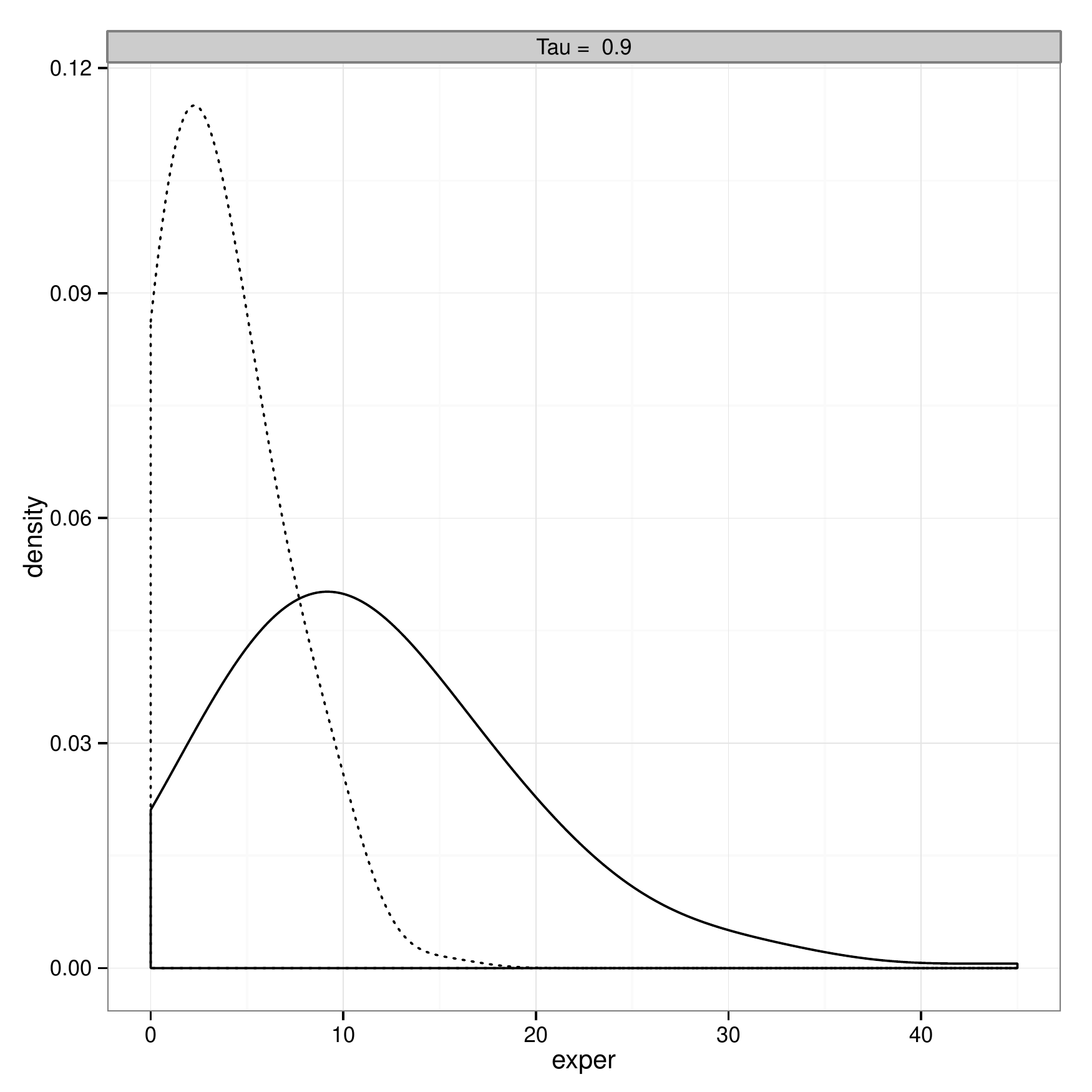}
\caption{Density of the variable {\it educ} separated in two groups according to their relative censoring probability in comparison with the mean probability in a given quantile, for $\tau = 0.5, 0.9$. Solid lines are for the group below the mean probability of being censored, while dashed lines are for the group above.}
\label{figure6}
  \end{center}
\end{figure}

Moreover, an analogous result is obtained with the variable years of experience and it is depicted in Figure~\ref{figure6}, but now considering quantiles 0.5 and 0.9. Considering the probabilities of being censored for $\tau = 0.50$, there is not a distinguishable difference between the distribution of years of experience from the group with probabilities below the average probability of being censored in this quantile against the women above the mean probability. Additionally, for $\tau = 0.9$, where the mean probability of being censored is 0.53, the group of women with probability below this are less experienced in terms of years of experience in the labour market, since their distribution is concentrated under 20 years of experience. The distribution for the other group of women peaks around 10 years, but it has a sizable part over 20 years of experience.

\subsection{Durable goods expenditures in Brazil}

We present here a second illustration of our model using data about household expenditures in Brazil, from the ``Consumer Expenditure Survey'' made between 2008 and 2009, which is a national survey that interviewed more than 50000 households around the country and it is available, in portuguese, at \url{http://www.ibge.gov.br/home/estatistica/populacao/condicaodevida/pof/2008_2009_encaa/microdados.shtm}. Due to computational limitations, we select the sample from a specific state, Maranh\~ao, to preserve to some extent the complex sampling scheme and also because this state had some similarities to the whole country data. After making this selection, we have 2240 observations, from which 1062 had zero expenditure with durable goods in the period. We include as covariates, gender ($x_1$: 0 = male, 1 = female), race ($x_2$ 0: white, 1: non-white), age in years ($x_3$), years of education ($x_4$), indicator variable if the individual has credit card ($x_5$: 0 = yes, 1 = no). Similar to the previous application, we again use a logistic model to analyze the probability of being equal to zero, where we add all variables and an intercept term. Also, all variables are included to explain the continuous part of the model. We decided to model a transformed response variable, namely $\sqrt{Y}$, instead of $Y$ due to information gain we achieve with this transformation, as before using this transformation, in the analysis for greater quantiles, all zero observations were being considered censored. Although this characteristic is not completely lost, as we see in the results presented here, it is vastly improved after using the square root of the expenditures with durable goods. So the following models are considered 
\begin{align} \label{finalModel}
 \log \left( \frac{p_i}{1-p_i} \right) &= \gamma_0 + \gamma_1 x_{i1} + \gamma_2 x_{i2} + \gamma_3 x_{i3} + \gamma_4 x_{i4} + \gamma_5 x_{i5}, \\
 Q_{\sqrt{Y_i}}(\tau | x_i) &= \beta_0 + \beta_1 x_{i1} + \beta_2 x_{i2} + \beta_3 x_{i3} + \beta_4 x_{i4} + \beta_5 x_{i5}.
\end{align}

\begin{figure}
\begin{center}
\includegraphics[scale=0.35]{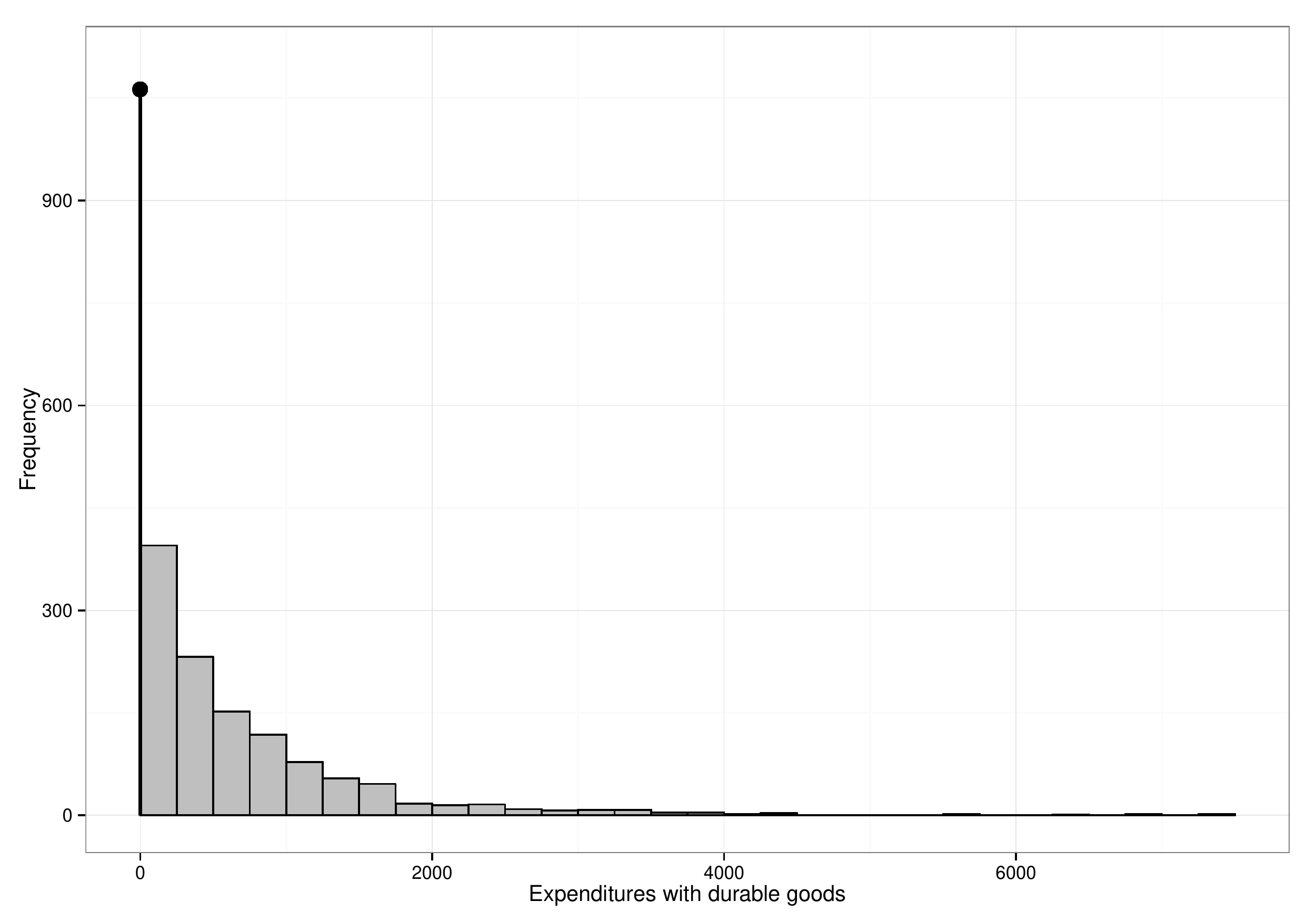}
\caption{Distribution of expenditures with durable goods, with a mass point at zero, in reais.}
\label{figureExp}
\end{center}
\end{figure}

\begin{table}
\centering
\caption{Mean posterior estimates and 90\% credible intervals for $\gamma$ in model~\ref{finalModel}, for $\tau  = 0.50$}\label{tableEst}
\begin{tabular}{lcc}
\hline
Variable & Estimate & Credible interval \\
Intercept & -2.14 & [-2.88 ; -1.49]\\
Gender & 0.02 & [-0.16 ; 0.18]\\
Race & 0.16 & [-0.05 ; 0.36]\\
Age & 0.08 & [-0.02 ; 0.18] \\
Education & -0.12 & [-0.23 ; -0.01]\\
Credit card & 0.79 & [0.50 ; 1.09]\\
\hline
\hline
\end{tabular}
\end{table}

We base our conclusions after running the MCMC obtaining 50000 samples from the posterior distribution of the parameters of interest, from which we discarded the first 10000 observations for burn-in purposes and later considered every 40th draw. As our estimator, we calculated the posterior mean for each parameter. Using these estimators, we see in Table~\ref{tableEst}, for the model of the probability $p_i$, that the only significant variables given their credible intervals are the indicator for credit card and years of education, where the former has a positive effect in the probability and the latter a negative effect in the probability of the expenditures being equal to zero, when $\tau = 0.5$. In comparison, we estimate the odds of having zero expenditures in a given period for a person who does not have a credit card is 2.2 times the odds of a person with a credit card. 

\begin{figure}
\begin{center}
\includegraphics[scale=0.35]{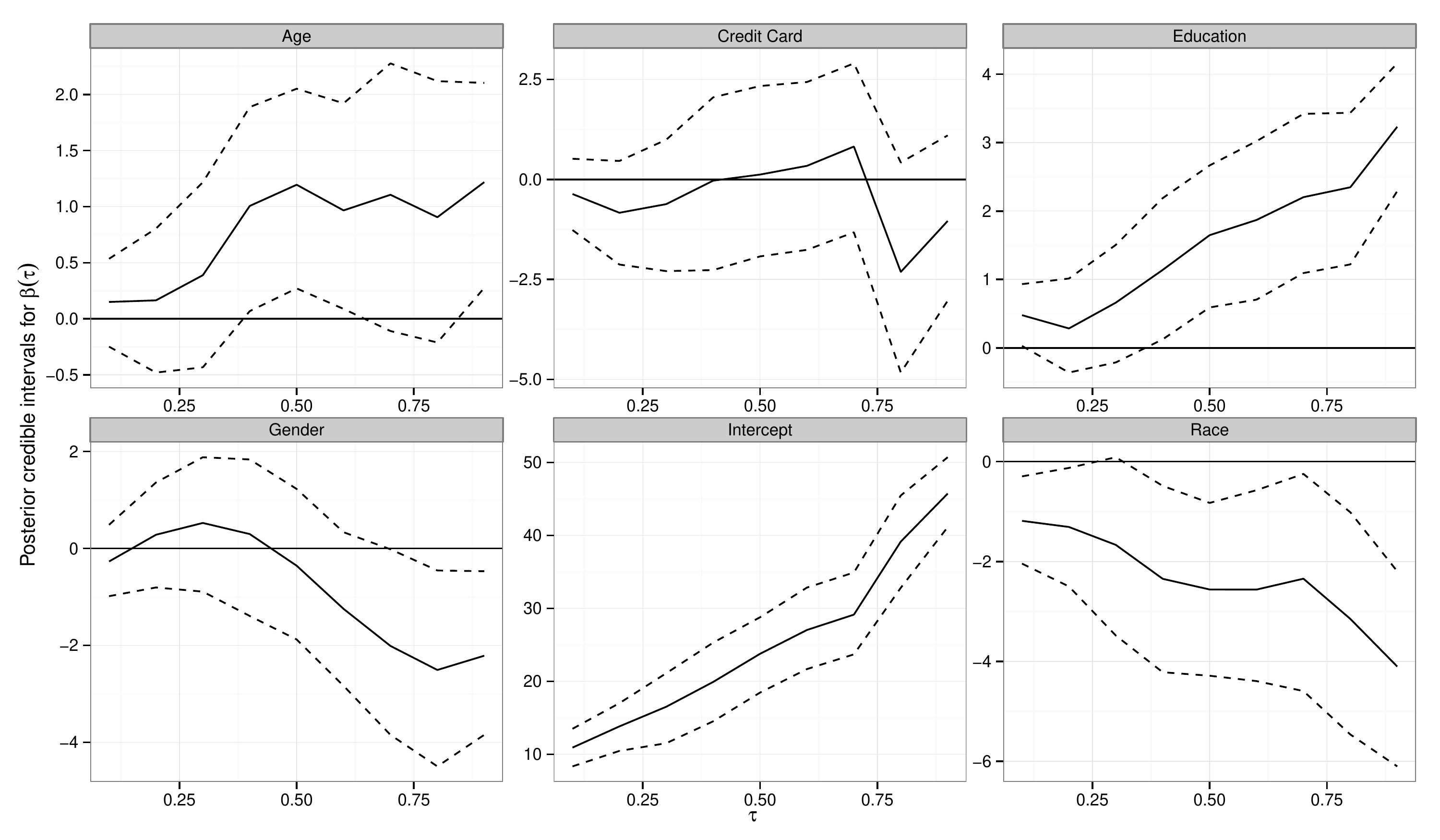}
\caption{Mean posterior estimates and 90\% credible intervals for $\beta(\tau)$}
\label{figurePostEst}
\end{center}
\end{figure}

The posterior mean estimates for the continuous part are shown in Figure~\ref{figurePostEst}. Using quantile regression models, we are able to check if some variables have significant effects in just some parts of the conditional distribution of the response variable. In this example, we find that difference between expenditures of men and women is only significant in the upper tail, or for $\tau = 0.8$ and $\tau = 0.9$, with a negative coefficient in this case. Similarly, the estimates for years of education and race are only significant for $\tau > 0.3$, where education has a positive effect in the expenditure, and race is estimated to have a negative effect. 

\begin{figure}
\begin{center}
\subfigure[]{
\resizebox*{5.5cm}{!}{\includegraphics{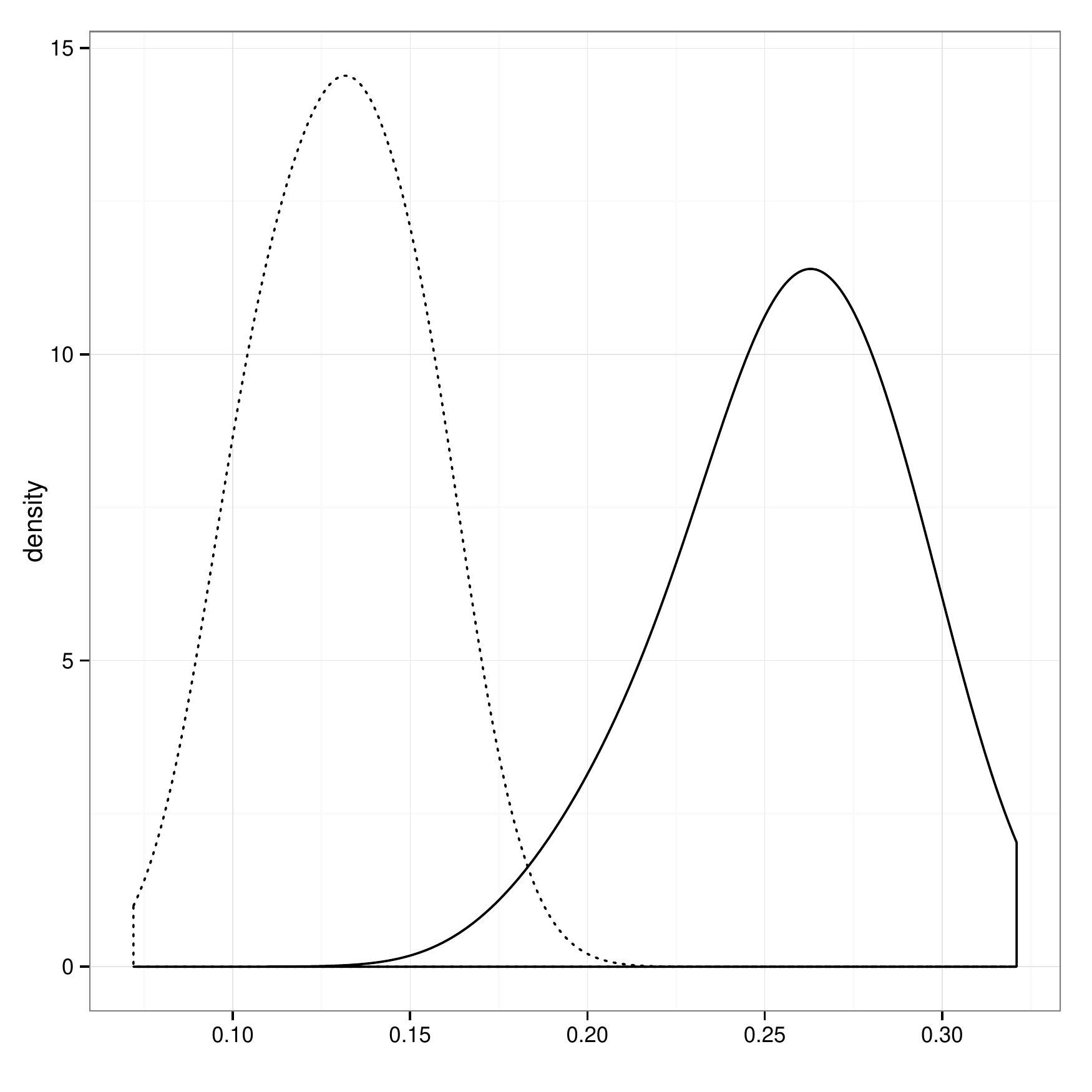}}
\label{fig10a}}\hspace{5pt}
\subfigure[]{
\resizebox*{5.5cm}{!}{\includegraphics{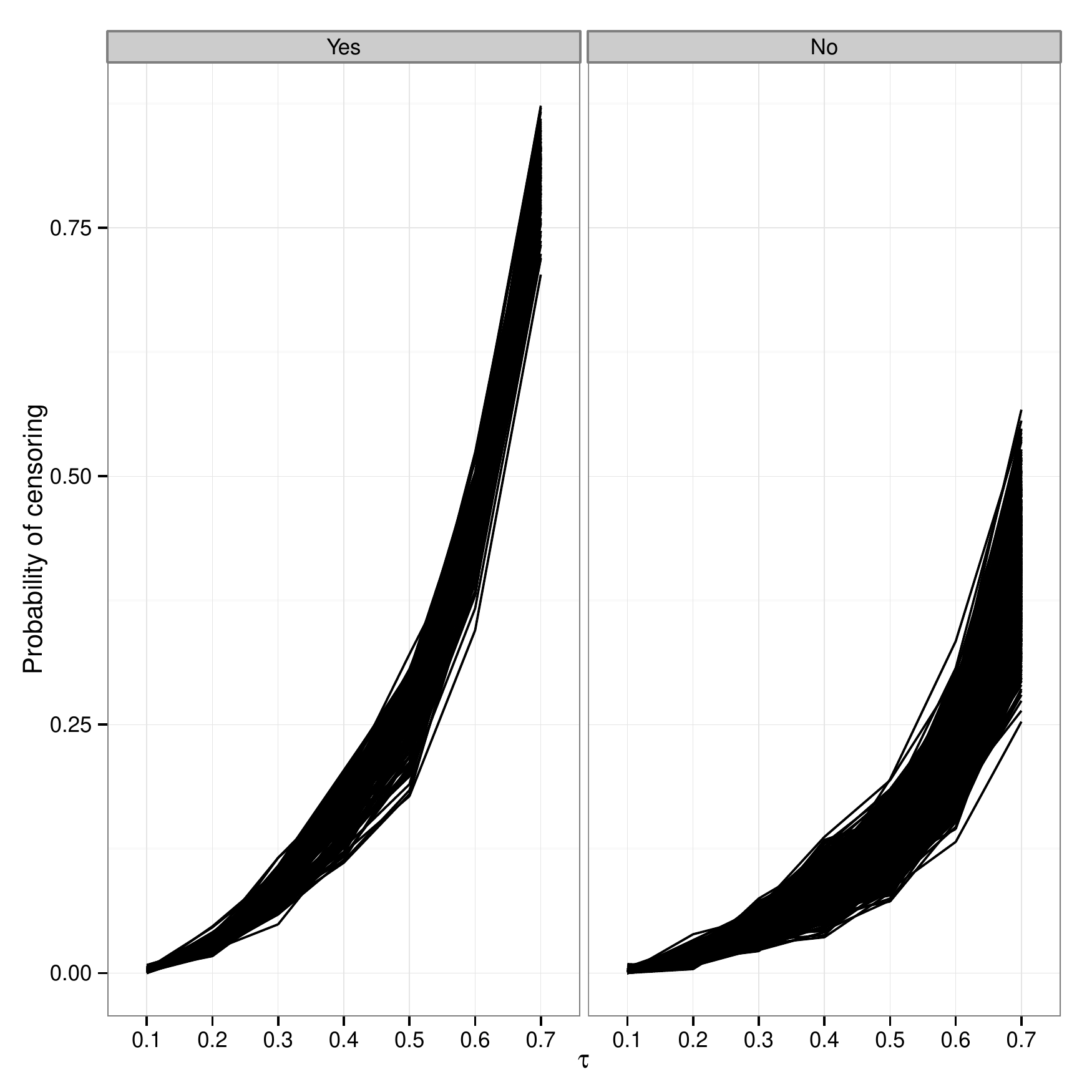}}
\label{fig10b}}
\caption{Comparisons for the probabilities of being censored given the indicator variable Credit Card (a) Estimated densities for $\tau = 0.5$, yes = solid line, no = dashed line. (b) Profiles of the probabilities for $\tau = 0.1, 0.2, \ldots, 0.7$.}
\label{figureCompProb}
\end{center}
\end{figure}

Analogous to the previous application, we can also compare the probability of being censored for different predictor variables. For $\tau = 0.5$, in Figure~\ref{fig10a} we see that the probability of being censored for people with credit cards is definitely greater than for people without one. This is in fact in concordance with the estimates in Table~\ref{tableEst}. Furthermore, we are able to check this information for other quantiles as well. As we notice in Figure~\ref{fig10b}, where we plot the probabilities of being censored for all observations with response value equal to zero, varying $\tau$ from 0.1 to 0.7, we recognize that, in general, the probabilities for the group with a credit card has a much faster increase when we move to greater quantiles. 

We do not show the probabilities for $\tau = 0.8$ and $\tau = 0.9$, because that was making it more difficult the comparison between those groups, since for these quantiles most observations are considered censored when their response is zero. This happens due to weight $F(0)$ has in the calculus in the probability of being censored. This factor in the probability depends on the scale of the data and that is the reason we used the square root transformation before starting the modeling process. We remember that in the previous application, there was no need for a transformation. We should note that for quantile regression models, one can use the equivariance to monotone transformations property for the quantile function, in order to get the estimated quantiles in the original scale of the data. But for the two-part model there is one problem with this particular setting, where most observations are considered censored for some quantiles, is that the posterior learning for the parameters $\gamma$ is compromised when this happens. In this example, we tested other transformations, but we left the square root just to emphasize that this feature should be carefully analyzed in each application.

\section{Final remarks}

In this paper, we extended the Bayesian Tobit quantile regression model to include the probability of a response variable equal to zero being censored, instead of considering them censored observations from the beginning, using the asymmetric Laplace distribution in the likelihood to analyze the conditional quantiles of the continuous part of the model. We used a two-part model to study the probability of $Y = 0$, denoting a point mass distribution at zero, while also building a linear predictor for the conditional quantiles of the continuous distributions for the response variable, meaning all observations where $Y > 0$ and when $Y$ is considered to be censored. 

We showed how the probability of being censored, in our model, is dependent on the quantile of interest, providing more information about whether one observation should be considered indeed censored, for instance, comparing the profiles of probabilities from different observations and checking their variation for smaller or greater $\tau$'s. We illustrated our findings with a well known dataset in the econometrics literature about female labour supply, where we showed how this probability of being censored given some covariates affects the model for different $\tau$'s. We also exemplified our model considering the problem of analyzing expenditures with durable goods in Brazil. Again, we demonstrate interesting results for the probabilities of being censored given the indicator variable of credit card. It is important to mention that our model could also be used in the survival analysis framework, when there is an assumption of cure in the study. Minor modifications would be necessary just to change the left censoring at zero for right censoring in this case. For future research, we are currently developing variable selection methods that try to share the information across both parts of the model.

\section*{Acknowledgements}

This research was supported by the Funda\c{c}\~ao de Amparo \`a Pesquisa do Estado de S\~ao Paulo (FAPESP) under Grants 2012/20267-9 and 2013/04419-6.

\bibliographystyle{imsart-nameyear}
\bibliography{Draft}

\end{document}